\documentclass{article}
\usepackage{latexsym}
\usepackage{epsfig}
\usepackage{amsmath}
\usepackage{amssymb}
\usepackage{amsthm}
\newtheorem{theorem}{Theorem}[section]
\newtheorem{corollary}[theorem]{Corollary}

\newtheorem{prop}[theorem]{Proposition}
\newtheorem{lemma}[theorem]{Lemma}

\theoremstyle{remark}
\newtheorem{remark}[theorem]{Remark}

\theoremstyle{definition}
\newtheorem{definition}[theorem]{Definition}

\begin{document}

\title{Separable balls around the maximally mixed state \\ for a 3-qubit system}

\author{Roland Hildebrand \thanks{%
LMC, Universit\'e Joseph Fourier, Tour IRMA, 51 rue des
Math\'ematiques, 38400 St.\ Martin d'H\`eres, France ({\tt
roland.hildebrand@imag.fr}).}}

\maketitle

\begin{abstract}
We obtain a new lower bound on the radius of the largest ball of separable unnormalized states around the
identity matrix for a 3-qubit system. This also enables us to improve the corresponding lower bounds for
multi-qubit systems. These bounds are approximately 5\% better than the previously known ones. As a
by-product, we compute the radius of the largest ball that fits into the triple projective tensor product of
the unit ball in $\mathbb R^3$.
\end{abstract}

\section{Introduction}

This work deals with lower bounds on the largest ball of separable unnormalized states around the identity matrix for a multi-partite quantum system
consisting of qubits. For a detailed introduction and a motivation of the paper we refer to \cite{Hildebrand0503194}.

In the latest version of \cite{Gurvits0409095} L.\ Gurvits and H.\ Barnum derived the bounds
obtained in \cite{Hildebrand0503194} by different means than those exploited in \cite{Hildebrand0503194}. Their ideas are applicable also in the case of systems
which are more complex than multi-qubit systems. In this contribution we improve these bounds, but again our result is limited to multi-qubit systems.
The result is based on the one hand on the ideas in \cite{Hildebrand0503194}, which apply in general to the multi-qubit case, and on the other hand on an idea
that is specific to the 3-qubit case. The latter will be the main subject of the present paper. 
As a by-product, we describe those points on the boundary of the projective tensor product (see \cite{Szarek0503221})
of 3 unit balls in $\mathbb R^3$ which are closest to the origin (Theorems \ref{PTP3},\ref{PPT3form}).

We prove that for a 3-qubit system a ball of radius $\sqrt{16/19}$ around the identity matrix consists only
of separable elements (Theorem \ref{3qubit}), as opposed to the best bound $\sqrt{4/5}$ known previously
\cite{Gurvits0409095},\cite{Hildebrand0503194}. For systems consisting of more than 3 qubits we obtain an
improvement of more than 5\% with respect to the best bounds known before \cite{Gurvits0409095}. Namely, we
prove that for an $m$-qubit system, a ball of radius $\frac{2^{m/2}}{\sqrt{\frac{17}{2}3^{m-3}+1}}$ around
the identity matrix consists only of separable elements (Theorem \ref{multiqubit}).

\smallskip

The paper is organized as follows.
In the next section we precisely define second order cones, positive and separable cones and tensor products
of more than two cones. Further we provide some simple properties of these objects and state some results
from \cite{Hildebrand0503194}, which will be needed later on. In Section 3 we consider the cone of separable
states for a system of 3 qubits and the related projective tensor product of unit balls in $\mathbb R^3$. In the
last section we apply the results obtained for 3 qubits to study the cone of separable unnormalized states of
a multi-qubit system.

\section{Definitions and preliminaries}

In this section we define the Lorentz cone, the cones of positive maps and the separable cones and provide some simple properties of these objects.
Further, we show how these cones are related to the cone of unnormalized separable states for a multi-qubit system.

\subsection{Separable elements and positive maps}

Denote the standard basis elements of $\mathbb R^n$ by $e_0^n,\dots,e_{n-1}^n$.
Tensor products of real vector spaces will have multi-indexed coordinates and basis elements. In particular, $e_{kl}$, $k=0,\dots,n-1$; $l=0,\dots,m-1$ will be the canonical basis elements of the space
$\mathbb R^n \otimes \mathbb R^m$. We identify this space with the space of real $n \times m$ matrices. An element $x \in \mathbb R^n \otimes \mathbb R^m$ will have coordinates $x_{kl}$,
$k=0,\dots,n-1$; $l=0,\dots,m-1$, corresponding to the entries of the representing matrix.
In this representation a product element $x \otimes y \in \mathbb R^n \otimes \mathbb R^m$ is
given by the rank 1 matrix $xy^T$.
Tensor products of three real vector spaces will have basis elements $e_{jkl}$ with three indices, and a point $x$ from such a tensor product space will have coordinates $x_{jkl}$.
The coordinates of a product $x = a \otimes b \otimes c$ will then be given by $x_{jkl} = a_jb_kc_l$.

{\definition Let $K,K'$ be regular convex cones, residing in finite-dimensional real vector spaces $E,E'$. We call a linear map $M: E \to E'$ {\it $K$-to-$K'$ positive} if $M[K] \subset K'$.
The set of $K$-to-$K'$ positive maps forms a regular convex cone, the {\it $K$-to-$K'$ positive cone}.}

{\definition Let $K_1,\dots,K_n$ be regular convex cones, residing in finite-dimensional real
vector spaces $E_1,\dots,E_n$. Then an element $w \in E_1 \otimes \dots \otimes E_n$ of the tensor product space is called {\it $K_1\otimes \dots \otimes K_n$-separable}
if it can be represented as a finite sum $\sum_{k=1}^N v_1^k \otimes \dots \otimes v_n^k$ of product elements
such that $v_l^k \in K_l$ for all $k = 1,\dots,N$; $l = 1,\dots,n$. The $K_1\otimes \dots \otimes K_n$-separable elements form a regular convex cone, the {\it $K_1\otimes \dots \otimes K_n$-separable cone}.
We denote this cone by $K_1\otimes \dots \otimes K_n$. }

\smallskip

It is easily seen that $K_1\otimes \dots \otimes K_n = K_1\otimes (K_2 \dots \otimes K_n) = (K_1\otimes \dots \otimes K_{n-1}) \otimes K_n$.

\medskip

We represent linear maps from ${\mathbb R}^m$ to ${\mathbb R}^n$ by $n \times m$ matrices.
Any linear map $M: \mathbb R^m \to \mathbb R^n$ can be viewed as linear functional on the space $\mathbb R^n \otimes \mathbb R^m$. On the product elements $x \otimes y \in \mathbb R^n \otimes \mathbb R^m$
it is defined as $\langle M, x \otimes y \rangle = \langle M(y), x \rangle$, where the scalar product on the right hand side is the Euclidean scalar product in $\mathbb R^n$. The vector spaces underlying the cone
of positive maps and the separable cone are hence dual to each other.
With the above matrix representation of the elements of $\mathbb R^n \otimes \mathbb R^m$ the scalar product
of a linear map $M: \mathbb R^m \to \mathbb R^n$ with an element $S \in \mathbb R^n \otimes \mathbb R^m$ is given by
$\langle M,S \rangle = tr\,(M^TS) = tr\,(S^TM)$. Here $^T$ denotes the transpose.
In general we identify the tensor product space ${\mathbb R}^{n_1} \otimes \dots \otimes {\mathbb R}^{n_m}$ with its dual by means of the Euclidean scalar product
$\langle x,y \rangle = \sum_{k_1=0}^{n_1-1} \dots \sum_{k_m=0}^{n_m-1} x_{k_1\dots k_m}y_{k_1\dots k_m}$.

{\remark Let $E$ be a real vector space equipped with a scalar product $\langle \cdot,\cdot \rangle$ and let $K \subset E$ be a convex cone. Then the dual cone $K^*$ is defined as
the set of elements $y \in E$ such that $\langle x,y \rangle \geq 0$ for all $x \in K$.}

\smallskip

We have the following standard result in the theory of positive maps.

{\theorem \label{dualth} Let $K,K'$ be regular convex cones. Then the $K \otimes K'$-separable cone is dual to the cone of
$K'$-to-$K^*$ positive maps with respect to the scalar product defined above.}

\begin{proof}
We have to show that the $K'$-to-$K^*$ positive maps are precisely those maps which are non-negative on all $K \otimes K'$-separable elements. By definition of the separable cone it suffices
to consider separable product elements $v \otimes v'$ with $v \in K$, $v' \in K'$.

Let $M$ be a positive map and let $v \in K$, $v' \in K'$. Then by definition $\langle M, v \otimes v' \rangle = \langle M(v'), v \rangle$. But $v' \in K'$, and $M$ is $K'$-to-$K^*$ positive. Hence
$M(v') \in K^*$. Since $v \in K$, we then have $\langle M(v'), v \rangle \geq 0$ by definition of the dual cone $K^*$. Therefore any positive map is in the dual cone to the $K \otimes K'$-separable cone.

Let now $M$ be in the dual cone to the $K \otimes K'$-separable cone. Then $\langle M(v'), v \rangle \geq 0$ for all $v \in K$, $v' \in K'$. But this is equivalent to $M(v') \in K^*$ for all $v' \in K'$, hence
$M$ is a $K'$-to-$K^*$ positive map.
\end{proof}

The dual cone $(K_1 \otimes K_2 \otimes K_3)^*$ to the $K_1 \otimes K_2 \otimes K_3$-separable cone is hence the cone of $K_3$-to-$P$ positive maps, where $P$ is the cone of
$K_2$-to-$K_1^*$ positive maps.

\medskip

Let $A: {\mathbb R}^n \to {\mathbb R}^n$ be a non-degenerate linear map, represented by an $n \times n$ matrix. We have the following result.

{\lemma \label{dualmap} Let $K \subset {\mathbb R}^n$ be a regular convex cone. Then the dual cone to the image $A[K]$ is the image $A^{-T}[K^*]$. }

\begin{proof}
Let $y \in {\mathbb R}^n$ be an element of the dual space to ${\mathbb R}^n$. Then $y \in (A[K])^*$ if and only if $\langle y, Ax \rangle = y^TAx \geq 0$ for all $x \in K$. This is in turn equivalent
to the condition $x^TA^Ty = \langle x, A^Ty \rangle \geq 0$ for all $x \in K$, which amounts to the inclusion $A^Ty \in K^*$. But this inclusion is equivalent to the inclusion $y \in A^{-T}[K^*]$.
\end{proof}

{\definition Let $E_1,\dots,E_n$ be real vector spaces and let $A_k: E_k \to E_k$, $k = 1,\dots,n$ be linear maps. Then the tensor product $A_1 \otimes \dots \otimes A_n$ of these maps is the unique linear map
$A : E_1 \otimes \dots \otimes E_n \to E_1 \otimes \dots \otimes E_n$ such that for any product element $x_1 \otimes \dots \otimes x_n \in E_1 \otimes \dots \otimes E_n$ we have
$A(x_1 \otimes \dots \otimes x_n) = A_1(x_1) \otimes \dots \otimes A_n(x_n)$. }

We have the following simple property of tensor products of maps.

{\lemma \label{tensormap}
Let $K_1 \subset E_1,\dots,K_n \subset E_n$ be regular convex cones and let $A_k: E_k \to E_k$, $k = 1,\dots,n$ be linear maps. Then with $A = A_1 \otimes \dots \otimes A_n$ we have the relation
$A[K_1\otimes \dots \otimes K_n] = A_1[K_1] \otimes \dots \otimes A_n[K_n]$. \qed }

\medskip

Let us introduce the following notion from \cite{Szarek0503221}.

{\definition Let $K_1 \subset E_1,\dots,K_n \subset E_n$ be convex bodies residing in finite-dimensional real vector spaces. Then their {\it projective tensor product}
$K = K_1 \otimes \dots \otimes K_n \subset E_1 \otimes \dots E_n$ is defined as the convex hull of the set $\{ x = x_1 \otimes \dots \otimes x_n \,|\, x_1 \in K_1,\dots,x_n \in K_n \}$. }

{\definition Let $K \in {\mathbb R}^n$ be a set. Then the {\it polar} $K^o$ of $K$ is defined to be the set $\{ x \in {\mathbb R}^n \,|\, |\langle x,y \rangle| \leq 1 \}$ for all $y \in K$. }

\smallskip

Let $\partial K$ denote the boundary of $K$. The following result follows easily from these definitions.

{\lemma \label{polarrad} Let $K$ be convex, compact and centrally symmetric and define $r_K = \min_{x \in \partial K} |x|$. Then $(K^o)^o = K$, $K^o$ is convex, compact and centrally symmetric and 
$\max_{x \in K^o} |x| = r_K^{-1}$. Moreover, if $y$ is a unit vector, then $r_K \cdot y \in \partial K$ if and only if $r_K^{-1} y \in K^o$. \qed }

\subsection{Lorentz cones and separable matrices}

Let $L_n$ be the $n$-dimensional standard Lorentz cone, or second order cone,
\[ L_n = \left\{ (x_0,\dots,x_{n-1})^T \in \mathbb R^n \,|\, x_0 \geq \sqrt{x_1^2 + \dots + x_{n-1}^2} \right\}.
\]
The Lorentz cone $L_n$ is {\it self-dual}, i.e.\ coincides with its dual cone with respect to the standard Euclidean metric in $\mathbb R^n$.
By the self-duality of the Lorentz cones a map $M: {\mathbb R}^m \to {\mathbb R}^n$ is $L_m$-to-$L_n$ positive if and only if $M^T$ is $L_m$-to-$L_n$ positive.
The central ray of $L_n$ is generated by the element $e_0^n$.

We will denote the $L_n$-to-$L_n$ positive cone by $P_n$. Then the dual cone to the triple tensor product $L_n \otimes L_n \otimes L_n$ will be the cone of $L_n$-to-$P_n$ positive maps.
Therefore an element $x$ is in $(L_n^{\otimes 3})^*$ if and only if the $n \times n$ matrix $M(y)$ defined elementwise by $M_{ij}(y) = x_{ij0} + \sum_{k=1}^{n-1} x_{ijk} y_k$ represents a $L_n$-to-$L_n$
positive map for all vectors $y \in \mathbb R^{n-1}$ of norm 1. Clearly for any $x \in (L_n^{\otimes 3})^*$ we have $x_{000} \geq 0$, and $x_{000} = 0$ implies $x = 0$.

\medskip

Let $I_k$ be the $k \times k$ identity matrix.
Denote the space of $n \times n$ hermitian matrices by ${\cal H}(n)$ and the cone of positive semidefinite matrices in this space by $H_+(n)$. Then we can identify the tensor space
$({\cal H}(n))^{\otimes m}$ with the space ${\cal H}(n^m)$. The tensor product of elements in ${\cal H}(n)$ amounts to the Kronecker product of matrices.
The set of unnormalized density matrices for an $n$-state quantum system is given by the cone $H_+(n)$. A quantum system consisting of $m$ subsystems with $n$ states each has
$n^m$ states, and the set of its unnormalized density matrices is $H_+(n^m)$. The set of unnormalized {\it separable} density matrices is a subset of $H_+(n^m)$, namely the tensor product
$(H_+(n))^{\otimes m}$.

We will concentrate on the case of multi-qubit systems, i.e.\ the case $n = 2$. Note that the space ${\cal H}(2)$ can be identified with ${\mathbb R}^4$ via the
orthonormal basis of normalized Pauli matrices (and the normalized identity matrix $I_2/\sqrt{2}$). Then the cone $H_+(2)$ will be identified with the Lorentz cone $L_4$ and the cone of separable unnormalized
density matrices of an $m$-qubit system with the cone $L_4^{\otimes m}$. Moreover, the identity matrix $I_2 \in {\cal H}(2)$ will be identified with the vector $\sqrt{2}e_0^4 \in {\mathbb R}^4$, and the
identity matrix $I_{2^m} = I_2^{\otimes m} \in {\cal H}(2^m)$ with the vector $2^{m/2} e_{0\dots0} \in ({\mathbb R}^4)^{\otimes m}$.

We are interested in the largest radius of a ball around $I_{2^m}$ that consists entirely of separable density matrices. It is equal to the largest radius of a ball around $2^{m/2} e_{0\dots0}$
that is contained in $L_4^{\otimes m}$.

\subsection{Linear images and largest radii of separable balls}

In this subsection we state results from \cite{Hildebrand0503194} that will be needed further.

\medskip

In the sequel we need linear images of Lorentz cones. The Lorentz cone $L_n$ can be represented as the conic hull of a closed ball in $\mathbb R^n$
which does not contain the origin. Therefore any linear image of $L_n$ can be represented as the conic hull of a closed ellipsoid not containing the origin, and any such conic hull is the linear image
of a Lorentz cone. We refer to such cones as {\it ellipsoidal cones}. Moreover, by a rotation of $\mathbb R^n$ such an image can be transformed to some standardized ellipsoidal cone.

{\definition Let $P$ be a positive definite real symmetric $(n-1)\times(n-1)$-matrix. Then the {\it standardized ellipsoidal cone} generated by $P$ will be
\[ K_{ell}(P) = \left\{ (x_0,x_1,\dots,x_{n-1})^T \,\big|\, x_0 \geq \sqrt{x^T P x},\ x = (x_1,\dots,x_{n-1})^T \right\} \subset \mathbb R^n.
\] }

It is not hard to establish the following result.

{\lemma \label{dualell} Let $P$ be a positive definite real symmetric $(n-1)\times(n-1)$-matrix. Then
$(K_{ell}(P))^* = K_{ell}(P^{-1})$. \qed }

\smallskip

The set of positive definite symmetric $(n-1)\times(n-1)$-matrices parameterizes the set of standardized ellipsoidal cones in $\mathbb R^n$.
We will need only ellipsoidal cones of a special type, namely conic hulls of balls with radius $\rho < 1$ around the element $e_0^n$. These cones are already in the standardized form defined above, with the matrix
$P$ being a multiple of the identity. The relation between $P$ and $\rho$ is given by the following lemma.

{\lemma \cite[Lemma 8]{Hildebrand0503194} \label{radii_rel} The cone generated by a ball of radius
$\rho < 1$ around the unit vector $e_0^n$ equals the standardized ellipsoidal cone $K_{ell}(r^{-2}I_{n-1}) \subset \mathbb R^n$ with
$r = \frac{\rho}{\sqrt{1-\rho^2}}$. }

Let us denote the cone $K_{ell}(r^{-2}I_{n-1}) \subset \mathbb R^n$ by $K_{ball}^n(r)$. Then $L_n = K_{ball}^n(1)$.
Identify ${\mathbb R}^{nm}$ with
${\mathbb R}^m\otimes {\mathbb R}^n$ by identifying the basis vectors $e_{kn+l}^{mn}$, $k = 0,\dots,m-1$; $l =
0,\dots,n-1$, with the orthonormal basis of tensor products $e_k^m \otimes e_l^n$. Then the cone
$K_{ball}^{nm}(r)$ is generated by a ball centered on $e_0^m \otimes e_0^n$.
From \cite{Hildebrand0503194} we have the following results.

{\lemma \cite[Corollary 4]{Hildebrand0503194} \label{ballr} Let $n,m \in {\mathbb N}$, $r_1,r_2 > 0$. Let $r$ be the largest number such that the inclusion $K_{ball}^{nm}(r) \subset K_{ball}^{n}(r_1) \otimes K_{ball}^{m}(r_2)$ holds. Then
\[ r = \left[ \max\left\{ -1 + (1 + r_1^{-2})(1 + r_2^{-2}), (\min(n,m)-1) r_1^{-2} r_2^{-2} \right\} \right]^{-1/2}.
\] }

\vspace{-0.5cm}

{\corollary \cite[Corollary 6]{Hildebrand0503194} \label{matrixballs} Let $B_{r_1} \subset {\cal H}(m)$, $B_{r_2} \subset {\cal H}(n)$ be balls of radii $r_1 < \sqrt{m}$, $r_2 < \sqrt{n}$ around the corresponding
identities $I_m,I_n$ and let $K_1,K_2$ be the conic hulls of these balls.
The largest ball around $I_{nm} \in {\cal H}(mn) = {\cal H}(m)\otimes {\cal
H}(n)$ which is contained in the cone of $K_1 \otimes K_2$-separable matrices has radius
\[ r = \min\left( r_1r_2,
\frac{\sqrt{mn}r_1r_2}{\sqrt{(\min(m^2,n^2)-1)(m-r_1^2)(n-r_2^2) + r_1^2r_2^2}} \right). \ \Box
\] }

\section{Largest ball of separable 3-qubit states}

In this section we obtain a lower bound on the radius of the largest separable ball around the identity matrix for a 3-qubit system.
This is the main part of the present contribution.

{\lemma \label{L2} Let $A,B$ be real symmetric matrices and $C$ a real matrix, all of size $n \times n$. Let $\lambda_1,\dots,\lambda_n$ be the eigenvalues of $A$
and $\mu_1,\dots,\mu_n$ be the eigenvalues of $B$, both in decreasing order. Suppose further that
\[ \left( \begin{array}{cc} A & C \\ C^T & B \end{array} \right) \succeq 0.
\]
Then $|C|_2^2 \leq \sum_{k=1}^n \lambda_k\mu_k$. }

\begin{proof}
The matrix inequality is fulfilled if and only if $C$ can be represented in the form $C = A^{1/2}MB^{1/2}$, where $\sigma_{\max}(M) \leq 1$. Since the 2-norm is a convex function and
the extreme points of the set $\{ M \,|\, \sigma_{\max}(M) \leq 1\}$ are the orthogonal matrices, we have $|C|_2 \leq \max_{U\,|\,UU^T = I} |A^{1/2}UB^{1/2}|_2$ and
$|C|_2^2 \leq \max_{U\,|\,UU^T = I} \langle A, UBU^T \rangle$. The set of orthogonal matrices is a smooth manifold, and the stationary points of the function $\langle A, UBU^T \rangle$
on this manifold are those matrices $U$ that make $A$ and $UBU^T$ commute. Thus we find an orthogonal matrix $V$ such that $\tilde A = VAV^T$ and $\tilde B = VUBU^TV^T$ are both diagonal. But then
the diagonal elements of $\tilde A$ are the $\lambda_k$ and the diagonal elements of $\tilde B$ are the $\mu_k$. Moreover, we have $\langle A, UBU^T \rangle = \langle \tilde A,\tilde B \rangle$. We obtain
$|C|_2^2 \leq \max_{\sigma \in S_n} \sum_{k=1}^n \lambda_k \mu_{\sigma(k)} = \sum_{k=1}^n \lambda_k \mu_k$. Here $S_n$ is the set of permutations of $\{1,\dots,n\}$.
\end{proof}

{\remark The lemma can be extended to the case when $A,B$ are of different size. }

{\lemma \label{L1} Let $M$ be an $n \times n$ matrix with singular values $\sigma_1,\dots,\sigma_n$. Then the matrix
\[ \left( \begin{array}{cc} \alpha I_n & M \\ M^T & \beta I_n \end{array} \right)
\]
has spectrum $\frac{\alpha+\beta}{2}\pm \sqrt{\left( \frac{\alpha-\beta}{2} \right)^2+\sigma_1^2},\dots,\frac{\alpha+\beta}{2}\pm \sqrt{\left( \frac{\alpha-\beta}{2} \right)^2+\sigma_n^2}$. }

\begin{proof}
Let $M = UDV$ be the singular value decomposition of $M$. Then we have
\[ \left( \begin{array}{cc} U^T & 0 \\ 0 & V \end{array} \right) \left( \begin{array}{cc} \alpha I_n & M \\ M^T & \beta I_n \end{array} \right) \left( \begin{array}{cc} U & 0 \\ 0 & V^T \end{array} \right)
= \left( \begin{array}{cc} \alpha I_n & D \\ D & \beta I_n \end{array} \right).
\]
Conjugation by an orthogonal matrix does not change the spectrum, and the spectrum of the matrix on the right-hand side is exactly
$\frac{\alpha+\beta}{2}\pm \sqrt{\left( \frac{\alpha-\beta}{2} \right)^2+\sigma_1^2},\dots,\frac{\alpha+\beta}{2}\pm \sqrt{\left( \frac{\alpha-\beta}{2} \right)^2+\sigma_n^2}$.
\end{proof}

The following lemma is a well-known result in the theory of matrix polynomials, and follows from the spectral factorization theorem \cite{Yakubovitch70}.

{\lemma \label{factor} Let $A_0,A_1,A_2$ be real symmetric matrices. Then the condition $A_0 + \cos\varphi A_1 + \sin\varphi A_2 \succeq 0$ for all $\varphi \in [-\pi,\pi]$ is equivalent to the linear matrix inequality
\[ \exists\ X = -X^T: \quad \left( \begin{array}{cc} A_0+A_1 & A_2+X \\ A_2-X & A_0-A_1 \end{array} \right) \succeq 0.
\] }

{\lemma \label{M012} Let $v = (v_1,v_2) \in \mathbb R^2$, let $M_0,M_1,M_2$ be $n \times n$ matrices, such that $\sigma_{\max}(M_0+\cos\varphi M_1+\sin\varphi M_2) \leq 1+\cos\varphi v_1+\sin\varphi v_2$
for all $\varphi \in [-\pi,\pi]$. Then $|M_0|_2^2+|M_1|_2^2+|M_2|_2^2 \leq 2n - (1-|v|^2)(n \mod 2)$. }

\begin{proof}
Assume the conditions of the lemma. Then we have $|v| \leq 1$.
Let $\varphi_0$ be an angle such that $v_1 = |v|\cos\varphi_0$, $v_2 = |v|\sin\varphi_0$. Then $\cos(\varphi+\varphi_0) v_1+\sin(\varphi+\varphi_0) v_2 \equiv \cos\varphi |v|$ as trigonometric
polynomials in $\varphi$.
Define further matrices $M'_1,M'_2$ by the identity $\cos(\varphi+\varphi_0) M_1+\sin(\varphi+\varphi_0) M_2 \equiv \cos\varphi M'_1+\sin\varphi M'_2$, where the left-hand and the right-hand side
are also considered as trigonometric polynomials in $\varphi$. Then we have $|M_1|_2^2+|M_2|_2^2 = |M'_1|_2^2+|M'_2|_2^2$
and $\sigma_{\max}(M_0+\cos\varphi M'_1+\sin\varphi M'_2) \leq 1+\cos\varphi |v|$ for all $\varphi \in [-\pi,\pi]$. This is equivalent to the condition
\[ \left( \begin{array}{cc} (1+\cos\varphi |v|)I_n & M_0+\cos\varphi M'_1+\sin\varphi M'_2 \\ (M_0+\cos\varphi M'_1+\sin\varphi M'_2)^T & (1+\cos\varphi |v|)I_n \end{array} \right)
\succeq 0 \qquad \forall\ \varphi \in [-\pi,\pi].
\]
By Lemma \ref{factor} this in turn is equivalent to the condition
\[ \exists\ X_1 = -X_1^T,X_2 = -X_2^T,X: \quad \left( \begin{array}{cccc} (1+|v|)I_n & X_1 & M_0+M'_1 & M'_2+X \\ -X_1 & (1-|v|)I_n & M'_2-X & M_0-M'_1 \\
M_0^T+(M'_1)^T & (M'_2)^T-X^T & (1+|v|)I_n & X_2 \\ (M'_2)^T+X^T & M_0^T-(M'_1)^T & -X_2 & (1-|v|)I_n \end{array} \right) \succeq 0.
\]
In particular, we have the matrix inequalities
\begin{equation} \label{cornerineq}
\left( \begin{array}{cc} (1+|v|)I_n & X_1 \\ -X_1 & (1-|v|)I_n \end{array} \right) \succeq 0, \qquad \left( \begin{array}{cc} (1+|v|)I_n & X_2 \\ -X_2 & (1-|v|)I_n \end{array} \right) \succeq 0.
\end{equation}
Let $\lambda_1,\dots,\lambda_n$ be the singular values of $X_1$, and $\mu_1,\dots,\mu_n$ the singular values of $X_2$, both in decreasing order. Then we have by Lemmata \ref{L2}, \ref{L1}
\begin{eqnarray*}
\lefteqn{\left| \left( \begin{array}{cc} M_0+M'_1 & M'_2+X \\ M'_2-X & M_0-M'_1 \end{array} \right) \right|_2^2 = 2(|M_0|_2^2 + |M'_1|_2^2 + |M'_2|_2^2 + |X|_2^2)} \\
&\leq& \sum_{k=1}^n \left[\left(1+\sqrt{|v|^2+\lambda_k^2}\right)\left(1+\sqrt{|v|^2+\mu_k^2}\right) + \left(1-\sqrt{|v|^2+\lambda_k^2}\right)\left(1-\sqrt{|v|^2+\mu_k^2}\right)\right] \\
&=& 2 \sum_{k=1}^n \left(1 + \sqrt{|v|^2+\lambda_k^2}\sqrt{|v|^2+\mu_k^2}\right).
\end{eqnarray*}
By (\ref{cornerineq}) we have $\lambda_k,\mu_k \leq \sqrt{1-|v|^2}$.
If $n$ is even, then we get the bound
\[ |M_0|_2^2 + |M'_1|_2^2 + |M'_2|_2^2 + |X|_2^2 \leq 2n.
\]
If $n$ is odd, then we have $\lambda_n = \mu_n = 0$, because a skew-symmetric matrix of odd size is singular. For $k = 1,\dots,n-1$ we again have the bound $\lambda_k,\mu_k \leq \sqrt{1-|v|^2}$.
Hence we obtain
\[ |M_0|_2^2 + |M'_1|_2^2 + |M'_2|_2^2 + |X|_2^2 \leq 2n - 1 + |v|^2.
\]
The assertion of the lemma now readily follows.
\end{proof}

{\lemma Let $v \in \mathbb R^m$ be a column vector and let $M_0,M_1,\dots,M_m$ be $n \times n$ matrices such that for all row vectors $x = (x_1,\dots,x_m) \in \mathbb R^m$ on the unit sphere we have
$\sigma_{\max}(M_0+\sum_{k=1}^mx_kM_k) \leq 1+xv$. Suppose that $n$ is odd. Then $|M_0|_2^2+\sum_{k=1}^m|M_k|_2^2 \leq 2n - (1-|v|^2) + (m-2)(n-1)(1-|v|^2)$. }

\begin{proof}
Assume the conditions of the lemma. Then we have $|v| \leq 1$. Let $U$ be an orthogonal $m \times m$ matrix such that $Uv = |v|e_1^m$ and define matrices $M'_l = \sum_{k=1}^mU_{lk}M_k$ and the vector
$y = (y_1,\dots,y_m) = xU^T$. Then we have $\sum_{k=1}^m |M_k|_2^2 = \sum_{k=1}^m |M'_k|_2^2$ and the condition $\sigma_{\max}(M_0+\sum_{k=1}^mx_kM_k) \leq 1+xv$ is equivalent to
$\sigma_{\max}(M_0+\sum_{k=1}^my_kM'_k) \leq 1+y_1|v|$. Let us consider two cases.

If $|v| = 1$, then for $y = -(e_1^m)^T$ we have $1+y_1|v| = 0$ and $M_0 = M'_1$. Let now $k > 1$ and set $y = (-e_1^m\cos\varphi+e_k^m\sin\varphi)^T$. This yields the condition
$\sigma_{\max}(M_0+\frac{\sin\varphi}{1-\cos\varphi} M'_k) \leq 1$ for all $\varphi \in [-\pi,\pi] \setminus \{0\}$. But $\frac{\sin\varphi}{1-\cos\varphi}$ tends to infinity as $\varphi \to 0$, hence $M'_k = 0$.
On the other hand, for $y = (e_1^m)^T$ we obtain $\sigma_{\max}(2M_0) \leq 2$, and $|M_0|_2^2 \leq n$. It follows that $|M_0|_2^2+\sum_{k=1}^m|M'_k|_2^2 \leq 2n$, which completes the proof
for this case.

Let now $|v| < 1$.
Since $n$ is odd, we have $\det\, M = -\det(-M)$ for any $n \times n$ matrix. Therefore any 2-dimensional linear subspace of the space of $n \times n$ matrices has a 1-dimensional subspace consisting
of singular matrices. We hence find an orthogonal matrix $V$ of size $(m-1) \times (m-1)$ and $n \times n$ matrices $M''_2,\dots,M''_m$ such that $M''_{k+1} = \sum_{l=1}^{m-1} V_{kl}M'_{l+1}$ for all
$k = 1,\dots,m-1$ and $M''_3,\dots,M''_m$ are singular. 
Define further a vector $y'$ by $y'_1 = y_1$, $(y'_2,\dots,y'_m) = (y_2,\dots,y_m)V^T$.

Then we have $\sigma_{\max}(M_0+y'_1M'_1+\sum_{k=2}^my'_kM''_k) \leq 1+y'_1|v|$ for all $y'$ on the unit sphere.
Consider the following linear fractional automorphism of the sphere $S^{m-1}$.
\begin{eqnarray*}
z &=& \frac{y'(\sqrt{1-|v|^2} I_m + (1-\sqrt{1-|v|^2}) e_1^m(e_1^m)^T) + (e_1^m)^T|v|}{1+y'_1|v|}, \\
y' &=& \frac{z(\sqrt{1-|v|^2} I_m + (1-\sqrt{1-|v|^2}) e_1^m(e_1^m)^T) - (e_1^m)^T|v|}{1-z_1|v|}
\end{eqnarray*}
Then the condition $\sigma_{\max}(M_0+y'_1M'_1+\sum_{k=2}^my'_kM''_k) \leq 1+y'_1|v|$ is equivalent to the condition $\sigma_{\max}(B_0+\sum_{k=1}^mz_kB_k) \leq 1$, where
\[ B_0 = \frac{1}{1-|v|^2}(M_0-|v|M'_1),\ B_1 = \frac{1}{1-|v|^2}(M'_1-|v|M_0),\ B_k = \frac{1}{\sqrt{1-|v|^2}} M''_k,\ k > 1.
\]
We have in particular $\sigma_{\max}(B_0\pm B_k) \leq 1$ and hence $\sigma_{\max}B_k \leq 1$ for all $k$. Therefore $|B_k|_2^2 \leq n-1$ for $k > 2$, because at least one singular value of $B_k$, $k > 2$,
is zero. This yields the bound $|M''_k|_2^2 \leq (1-|v|^2)(n-1)$ for $k > 2$.
In addition, by the preceding lemma we have $|M_0|_2^2+|M'_1|_2^2+|M''_2|_2^2 \leq 2n - (1-|v|^2)$.

This finally yields $|M_0|_2^2+\sum_{k=1}^m|M_k|_2^2 = |M_0|_2^2+|M'_1|_2^2+\sum_{k=2}^m|M''_k|_2^2 \leq 2n - (1-|v|^2) + (m-2)(n-1)(1-|v|^2)$.
\end{proof}

{\corollary \label{sqrt7} Let $v \in \mathbb R^3$ be a column vector and let $M_0,M_1,M_2,M_3$ be $3 \times 3$ matrices such that $\sigma_{\max}(M_0+\sum_{k=1}^mx_kM_k) \leq 1+xv$ for all row vectors $x$ on the
unit sphere. Then $|M_0|_2^2 + |M_1|_2^2 + |M_2|_2^2 + |M_3|_2^2 + |v|^2 \leq 7$. \qed }

\smallskip

This bound is exact, as the example
\begin{equation} \label{matrixex}
M_0 = 0,\ v = 0,\ M_1 = \left( \begin{array}{ccc} 1 & 0 & 0 \\ 0 & -1 & 0 \\ 0 & 0 & -1 \end{array}
\right),\ M_2 = \left( \begin{array}{ccc} 0 & -1 & 0 \\ -1 & 0 & 0 \\ 0 & 0 & 0 \end{array} \right),\ M_3 =
\left( \begin{array}{ccc} 0 & 0 & -1 \\ 0 & 0 & 0 \\ -1 & 0 & 0 \end{array} \right)
\end{equation}
shows. This example together with the preceding corollary yields the following result.

{\theorem \label{PTP3} Let $B \subset {\mathbb R}^3$ be the unit ball. Then the largest ball centered on the
origin in $({\mathbb R}^3)^{\otimes 3}$ that is contained in the projective tensor product $B^{\otimes 3} = B \otimes B
\otimes B$ has radius $\sqrt{1/7}$. }

\begin{proof}
In this proof we will not deal with cones, but with bounded convex bodies. We hence adapt throughout the proof a numeration of the indices that commences with 1. The coordinates of a vector $x \in {\mathbb R}^3$ will hence be $x_1,x_2,x_3$.

Note that $B^{\otimes 3}$ is convex, compact, and centrally symmetric.
A point $w \in ({\mathbb R}^3)^{\otimes 3}$ is in the polar of $B^{\otimes 3}$ if and only if $\sum_{j,k,l = 1}^3 w_{jkl} x_j y_k z_l \leq 1$ for any triple
of unit length vectors $x,y,z \in {\mathbb R }^3$. Define $3 \times 3$ matrices $M_j$, $j = 1,2,3$, by $(M_j)_{kl} = w_{jkl}$. Then the condition $w \in (B^{\otimes 3})^o$ is equivalent to the condition
$\sigma_{\max}(\sum_{k=1}^3 x_jM_j) \leq 1$ for all $x \in B$. We have $|w|^2 = \sum_{j=1}^3 |M_j|_2^2$, hence by the preceding corollary and in view of example (\ref{matrixex}) we have
$\max_{w \in (B^{\otimes 3})^o} |w| = \sqrt{7}$. An application of Lemma \ref{polarrad} concludes the proof.
\end{proof}

The following result shows that the matrices $M_1,M_2,M_3$ in (\ref{matrixex}) define essentially the only element in $(B^{\otimes 3})^o$ which has norm $\sqrt{7}$.

{\theorem \label{PPT3form} Let $B \subset {\mathbb R}^3$ be the unit ball and let $x \in \partial B^{\otimes 3}$ be such that $|x| = \sqrt{1/7}$. Then there exist orthogonal $3 \times 3$ matrices $U,V,W$ such that
$x' = 7(U \otimes V \otimes W)x$ is given componentwise by $x'_{jkl} = (M_l)_{jk}$, $j,k,l = 1,2,3$, where the matrices $M_1,M_2,M_3$ are given by (\ref{matrixex}). }

\begin{proof}
By Lemma \ref{polarrad} we have only to show that any element $y \in (B^{\otimes 3})^o$ of norm $\sqrt{7}$ can be brought to the form defined by the matrices $M_1,M_2,M_3$ by a transformation of the form
$U \otimes V \otimes W$, where $U,V,W$ are orthogonal. 

Let $y \in (B^{\otimes 3})^o$, $|y|^2 = 7$ and define matrices $Y_1,Y_2,Y_3$ by $(Y_l)_{jk} = y_{jkl}$, $j,k,l = 1,2,3$. Assume without loss of generality that $Y_3$ is singular (otherwise we apply an appropriate
transformation of the form $Y_2 \mapsto \cos\varphi Y_2 + \sin\varphi Y_3$, $Y_3 \mapsto -\sin\varphi Y_2 + \cos\varphi Y_3$, which amounts to the application of a map $Id \otimes Id \otimes W$ to $y$,
where $W$ is a rotation in the $(2,3)$-plane).
Then $|Y_3|_2^2 \leq 2$ and by Lemma \ref{M012} $|Y_1|_2^2 + |Y_2|_2^2 \leq 5$. It follows that both inequalities are in fact equalities. In particular, the pendants of the matrices $M_0,X$ in the proof of Lemma
\ref{M012} must be zero, and the pendants of $X_1,X_2$ must have 2-norm $\sqrt{2}$. This leads to the following condition on $Y_1,Y_2$.
\[ \exists\ X_1 = -X_1^T,X_2 = -X_2^T: \quad \left( \begin{array}{cccc} I_3 & X_1 & Y_1 & Y_2 \\ -X_1 & I_3 & Y_2 & -Y_1 \\
Y_1^T & Y_2^T & I_3 & X_2 \\ Y_2^T & -Y_1^T & -X_2 & I_3 \end{array} \right) \succeq 0; \qquad |X_1|_2^2 = |X_2|_2^2 = 2.
\]
It can be written in a more compact form as
\[ \exists\ X_1 = -X_1^T,X_2 = -X_2^T: \quad \left( \begin{array}{cc} I_3+iX_1 & Y_1+iY_2 \\ Y_1^T-iY_2^T & I_3+iX_2 \end{array} \right) \succeq 0; \qquad |X_1|_2^2 = |X_2|_2^2 = 2.
\]
There exist orthogonal matrices $U,V$ such that $UX_1U^T$ and $VX_2V^T$ equal the matrix
\[ J = \left( \begin{array}{ccc} 0 & 1 & 0 \\ -1 & 0 & 0 \\ 0 & 0 & 0 \end{array} \right).
\]
By conjugating above matrix inequality with $diag(U,V)$, we get the matrix inequality
\[ \left( \begin{array}{cc} I_3+iJ & U(Y_1+iY_2)V^T \\ V(Y_1^T-iY_2^T)U^T & I_3+iJ \end{array} \right) \succeq 0.
\]
Then the relation $|U(Y_1+iY_2)V^T|_2^2 = 5$ can hold only if $UY_1V^T = I_3$, $UY_2V^T = J$.

Note that $\sigma_{\max}(\cos\varphi Y_1 + \sin\varphi Y_3) = \sigma_{\max}(\cos\varphi I_3 + \sin\varphi UY_3V^T) \leq 1$ for all $\varphi$. This is possible only if $UY_3V^T$ is skew-symmetric.
Moreover, we have $\sigma_{\max}(\cos\varphi Y_2 + \sin\varphi Y_3) = \sigma_{\max}(\cos\varphi UY_2V^T + \sin\varphi UY_3V^T) \leq 1$ for all $\varphi$. Since both matrices $UY_2V^T$ and $UY_3V^T$
are skew-symmetric, this is possible only if they are orthogonal to each other. But then we find another orthogonal matrix $U'$ such that
\[ U'UY_1V^T(U')^T = I_3,\quad
U'UY_2V^T(U')^T = \left( \begin{array}{ccc} 0 & 1 & 0 \\ -1 & 0 & 0 \\ 0 & 0 & 0 \end{array} \right),\quad
U'UY_3V^T(U')^T = \left( \begin{array}{ccc} 0 & 0 & 1 \\ 0 & 0 & 0 \\ -1 & 0 & 0 \end{array} \right).
\]
The theorem now follows by multiplying all these three products by the orthogonal matrix $diag(1,-1,-1)$ from the right.
\end{proof}

{\prop Let $x \in (L_4^{\otimes 3})^*$ with $x_{000} = 1$. Then the inequality $\sum_{k=1}^3 x_{00k}^2 + \sum_{k,l=1}^3 x_{kl0}^2 + \sum_{i,j,k = 1}^3 x_{ijk}^2 \leq 7$ holds. }

\begin{proof}
Let $x$ satisfy the assumptions of the proposition and define the mapping $D: {\mathbb R}^4 \to {\mathbb R}^4$ by $D = diag(1,-1,-1,-1)$. Note that the Lorentz cone $L_4$ is invariant with respect to this transformation.
Then by Lemma \ref{tensormap} the cone $L_4^{\otimes 3}$ is invariant with respect to the map $D \otimes D \otimes Id$. From Lemma \ref{dualmap} it therefore follows that $(L_4^{\otimes 3})^*$ is invariant
with respect to the map $(D \otimes D \otimes Id)^{-T} = D \otimes D \otimes Id$. Define a point $x' = (D \otimes D \otimes Id)(x)$. It is given coordinatewise by
\[ x'_{ijk} = \left\{ \begin{array}{rcl} x_{ijk}, & \ & i = j = 0, \\ x_{ijk}, & \ & i \not= 0,\  j \not= 0, \\ -x_{ijk}, & & ij = 0,\ i+j \not= 0, \end{array} \right.
\]
and is also in $(L_4^{\otimes 3})^*$. By convexity of this cone we have $y = \frac{x+x'}{2} \in (L_4^{\otimes 3})^*$. Note that $y_{i0k} = 0$ for all $i \not=0$ and all $k$; and
$y_{0jk} = 0$ for all $j \not= 0$ and all $k$. For all other elements we have $y_{ijk} = x_{ijk}$.

Let us now define the vector $v \in \mathbb R^3$ by $v_k = x_{00k}$, the $3 \times 3$ matrix $M_0$ by $(M_0)_{ij} = x_{ij0}$ and the $3 \times 3$ matrices $M_k$, $k=1,2,3$, by $(M_k)_{ij} = x_{ijk}$.
Then the condition $y \in (L_4^{\otimes 3})^*$ is equivalent to the condition
\[ \left( \begin{array}{cc} 1+zv & 0 \\ 0 & M_0+\sum_{k=1}^mz_kM_k \end{array} \right) \in P_4 \qquad \forall\ z = (z_1,z_2,z_3) \in {\mathbb R}^3,\ |z| = 1.
\]
This in turn is equivalent to the condition $\sigma_{\max}(M_0+\sum_{k=1}^mz_kM_k) \leq 1+zv$ for all row vectors $z \in \mathbb R^3$, $|z| = 1$.
Corollary \ref{sqrt7} then completes the proof.
\end{proof}

Consider the inequality $\sum_{k=1}^3 x_{00k}^2 + \sum_{k,l=1}^3 x_{kl0}^2 + \sum_{i,j,k = 1}^3 x_{ijk}^2 \leq 7$ from the preceding proposition.
If we permute the indices and combine all resulting inequalities, then we obtain the following corollary.

{\corollary \label{ungl1} Let $x \in (L_4^{\otimes 3})^*$ with $x_{000} = 1$. Then the inequality $\frac{1}{3}\sum_{k=1}^3 (x_{00k}^2+x_{0k0}^2+x_{k00}^2) +
\frac{1}{3}\sum_{k,l=1}^3 (x_{kl0}^2+x_{k0l}^2+x_{0kl}^2) + \sum_{i,j,k = 1}^3 x_{ijk}^2 \leq 7$ holds. \qed}

\smallskip

This means that $(L_4^{\otimes 3})^*$ is contained in the ellipsoidal cone defined by this inequality.

\medskip

We now make repeatedly use of Lemma \ref{ballr}. Consider the space $({\mathbb R}^4)^{\otimes 2} = {\mathbb
R}^{16}$ and the cone $L_4 \otimes L_4 = K_{ball}^4(1) \otimes K_{ball}^4(1)$ in this space. By Lemma
\ref{ballr} this cone contains the cone $K_{ball}^{16}(1/\sqrt{3})$. Again by Lemma \ref{ballr} the cone
$K_{ball}^{16}(1/\sqrt{3}) \otimes K_{ball}^4(\sqrt{5/3})$ contains the cone $K_{ball}^{64}(\sqrt{5/27})$.
Hence $K_{ball}^{64}(\sqrt{5/27}) \subset L_4 \otimes L_4 \otimes K_{ball}^4(\sqrt{5/3})$.

Now note that the cone $K_{ball}^4(\sqrt{5/3})$ is a linear image of $L_4$. Namely, if $D$ is a linear map
given by the diagonal matrix $diag(1,\sqrt{\frac{3}{5}},\sqrt{\frac{3}{5}},\sqrt{\frac{3}{5}})$, then
$D[K_{ball}^4(\sqrt{5/3})] = L_4$. Therefore $(Id \otimes Id \otimes D)[K_{ball}^{64}(\sqrt{5/27})] \subset
(Id \otimes Id \otimes D)[L_4 \otimes L_4 \otimes K_{ball}^4(\sqrt{5/3})] = L_4^{\otimes 3}$. Let us
formalize this result in the following lemma.

{\lemma Let $y \in ({\mathbb R}^4)^{\otimes 3}$ be such that $y_{000} = 1$ and the inequality $\sum_{k=1}^3
(y_{k00}^2+y_{0k0}^2) + \sum_{k,l=1}^3 y_{kl0}^2 + \frac{5}{3}\left( \sum_{k=1}^3 y_{00k}^2 + \sum_{k,l=1}^3
(y_{k0l}^2+y_{0kl}^2) + \sum_{i,j,k=1}^3 y_{ijk}^2 \right) \leq \frac{5}{27}$ holds. Then $y \in L_4^{\otimes
3}$. \qed }

\smallskip

By Lemma \ref{dualell} we obtain the following corollary.

{\corollary Let $x \in (L_4^{\otimes 3})^*$ with $x_{000} = 1$. Then the inequality $\sum_{k=1}^3 (x_{k00}^2+x_{0k0}^2) + \sum_{k,l=1}^3 x_{kl0}^2 + \frac{3}{5}\left( \sum_{k=1}^3 x_{00k}^2
+ \sum_{k,l=1}^3 (x_{k0l}^2+x_{0kl}^2) + \sum_{i,j,k=1}^3 x_{ijk}^2 \right) \leq \frac{27}{5}$ holds. \qed}

\smallskip

Again by a permutation of the indices and combining the inequalities we obtain the following corollary.

{\corollary Let $x \in (L_4^{\otimes 3})^*$ with $x_{000} = 1$. Then the inequality $13 \sum_{k=1}^3 (x_{k00}^2+x_{0k0}^2+x_{00k}^2) + 11 \sum_{k,l=1}^3 (x_{kl0}^2+x_{k0l}^2+x_{0kl}^2) + 9
\sum_{i,j,k=1}^3 x_{ijk}^2 \leq 81$ holds. \qed}

\smallskip

Adding 3 times the inequality from Corollary \ref{ungl1} we obtain the following theorem.

{\prop Let $x \in (L_4^{\otimes 3})^*$ with $x_{000} = 1$. Then the inequality $7 \sum_{k=1}^3 (x_{k00}^2+x_{0k0}^2+x_{00k}^2) + 6 \sum_{k,l=1}^3 (x_{kl0}^2+x_{k0l}^2+x_{0kl}^2) + 6
\sum_{i,j,k=1}^3 x_{ijk}^2 \leq 51$ holds. \qed}

{\corollary The inclusions $(L_4^{\otimes 3})^* \subset K_{ball}^{64}(\sqrt{17/2})$ and
$K_{ball}^{64}(\sqrt{2/17}) \subset L_4^{\otimes 3}$ hold. \qed}

\smallskip

Note that the conic hull of a ball of radius $\rho$ around $e_0^{64}$ and the conic hull of a ball of radius $l\rho$ around $le_0^{64}$, $l > 0$, are equal.
The following theorem is then a consequence of the preceding corollary, Lemma \ref{radii_rel} and the fact that the identity $I_8 \in {\cal
H}(8) \cong \mathbb R^{64}$ has norm $\sqrt{8}$.

{\theorem \label{3qubit} Let $B \subset {\cal H}(8) = ({\cal H}(2))^{\otimes 3}$ be an Euclidean ball of radius
$\sqrt{\frac{16}{19}}$ around the identity matrix $I_8$. Then $B$ consists of tripartite separable elements.
\qed }

\smallskip

The theorem yields an improved lower bound on the largest radius of a tripartite separable ball around the
identity matrix. The best previously known bound was $\sqrt{\frac{4}{5}}$, obtained independently by Leonid
Gurvits by a different line of reasoning and by the author by using iteratively Corollary \ref{matrixballs}
on a triple of Lorentz cones $L_4$.

\section{Largest ball of separable $n$-qubit states}

Applying Corollary \ref{matrixballs} further, one gets improved lower bounds on the largest radius of a ball
of separable multi-qubit states around the identity.

Following the line of reasoning in \cite[Section 5]{Hildebrand0503194}, we define a sequence $\rho_k$ recursively by
\[ \rho_3^{-2} = \frac{19}{16};\quad \rho_k^{-2} = \frac{3}{2}\rho_{k-1}^{-2} - 2^{-k+1},\ k>3.
\]
We have the explicit expression
\[ \rho_k^{-2} = \frac{17}{54}\left( \frac{3}{2} \right)^k + 2^{-k},\quad \rho_k =
\frac{2^{k/2}}{\sqrt{\frac{17}{2}3^{k-3}+1}},\quad k \geq 3.
\]

By the repeated use of Corollary \ref{matrixballs} we obtain the following theorem (see also \cite{Hildebrand0503194}).

{\theorem \label{multiqubit} For $k \geq 3$ the expression $\rho_k = \frac{2^{k/2}}{\sqrt{\frac{17}{2}3^{k-3}+1}}$ is a lower bound on the
radius of the largest separable ball of unnormalized multi-partite mixed states of a $k$-qubit system around
the identity matrix in the space ${\cal H}(2)^{\otimes k}$. \qed}

\smallskip

For $n$-qubit systems with $n \geq 3$ we get an improvement of over $5\%$ with respect to the results in
\cite{Gurvits0409095},\cite{Hildebrand0503194}.

\end{document}